\documentclass[aps,pre,twocolumn,showpacs,superscriptaddress,amsmath,amssymb]{revtex4-1}
\usepackage{epsfig}
\usepackage{amsmath}
\usepackage{MnSymbol}
\usepackage{siunitx}
\usepackage{soul}
\usepackage{array}
\usepackage{multirow}
\usepackage{psfrag}
\usepackage[font=footnotesize,format=plain,labelsep=period,justification=raggedright]{caption}
\usepackage{subcaption}
\usepackage{xcolor}
\usepackage{graphicx}

\definecolor{colorgaby}{RGB}{0 0 255}
\definecolor{rla}{RGB}{0 200 100}

\let\oldhat\hat

\renewcommand{\hat}[1]{\oldhat{\mathbf{#1}}}

\begin{document}
\title{Apparent Contact Angle of Droplets on Liquid Infused Surfaces: \\ Geometric Interpretation}
\author{Ciro Semprebon}
\affiliation{Smart Materials and Surfaces Laboratory, Northumbria University, Newcastle upon Tyne NE1 8ST, UK.}
\author{Muhammad Subkhi Sadullah}
\affiliation{King Abdullah University of Science and Technology (KAUST), Water Desalination and Reuse Center (WDRC), Biological and Environmental Science and Engineering (BESE) Division, Thuwal 23955-6900, Saudi Arabia.}
\author{Glen McHale}
\affiliation{School of Engineering, The University of Edinburgh, Kings Buildings, Edinburgh EH9 3FB, UK.}
\author{Halim Kusumaatmaja}
\affiliation{Department of Physics, Durham University, Durham, DH1 3LE, UK.}

\date{\today}

\begin{abstract}
We theoretically investigate the apparent contact angle of droplets on liquid infused surfaces as a function of the relative size of the wetting ridge and the deposited droplet. We provide an intuitive geometrical interpretation whereby the variation in the apparent contact angle is due to the rotation of the Neumann triangle. We also derive linear and quadratic corrections to the apparent contact angle as power series expansion in terms of pressure differences between the lubricant, droplet and gas phases. These expressions are much simpler and more compact compared to those previously derived by Semprebon et al. [Soft Matter, 2017, 13, 101-110].
\end{abstract}

\maketitle
\renewcommand{\arraystretch}{1.5}


\section{Introduction}

In recent years there have been expanding interest in a novel class of functional surfaces commonly termed as lubricant impregnated/liquid infused surfaces (LIS) or slippery liquid infused porous surfaces (SLIPS) \cite{Lafuma_2011,Wong2011,Smith_2013}. On LIS, a lubricant is trapped in between the surface textures by capillary forces, and its presence can result in numerous advantageous surface properties, such as self-cleaning, enhanced heat transfer, anti-fouling, anti-icing. LIS are also considered superior compared to other liquid repellent surfaces (e.g. classical superhydrophobic surfaces) because they are robust against low surface tension liquids and pressure-induced failures. These features make LIS potentially transformative for a wide range of applications \cite{Li_2019,Villegas_2019, Ware_2018, Preston_2018, Weisensee2017,Sadullah_2020_2}, including for marine and medical coatings, product packaging, heat exchanger, water harvesting and droplet microfluidics.

Fundamentally, the presence of the lubricant also distinguishes LIS from other wetting scenarios. In classical wetting phenomena, we typically consider three phases, corresponding to the solid, droplet and gas phases, and there is one three-phase contact line. In contrast, for an LIS system, there are four phases to consider: the solid, lubricant, droplet and gas phases; and there can be up to 3 distinct three-phase contact lines: the lubricant-gas-solid,
lubricant-droplet-solid, and lubricant-droplet-gas contact lines. These various three-phase contact lines are connected at the wetting ridge at the foot of the droplet, and this wetting ridge is key to understanding many static and dynamic behaviours of droplets on LIS. These range from the shape of a droplet on LIS \cite{Schellenberger2015,Tress_2017,Sadullah_2018,Gunjan_2020} and its adhesion to the substrate \cite{Shek_2021} to what dominates the pinning force \cite{Sadullah_2020} and viscous dissipation \cite{Keiser2017,Daniel2017,Kreder_2018,Keiser_2020} during droplet motion.

In this work, we will focus on the suitable measure of wettability on LIS, a key design parameter for any application involving LIS. Since the solid-droplet-gas contact line is not present on LIS, the standard Young's equation for contact angle cannot be employed. Instead, we recently proposed the notion of an apparent contact angle \cite{Guan_2015,Semprebon2016b}, and demonstrated its applicability both against simulation \cite{Sadullah_2018} and experimental \cite{Schellenberger2015} data. The simplest case for the apparent angle is in the limit where the wetting ridge size is negligible compared to the droplet. In this case, the apparent angle relation can be compactly written as a function of six independent surface tensions due to the presence of 4 separate phases for LIS, and the solid surface geometry. When the wetting ridge size is not negligible, the apparent angle is no longer uniquely defined by material parameters, but it is also a function of the shape and size of the wetting ridge. The previous expression derived by Semprebon et al. \cite{Semprebon2016b}, however, was complex, and more importantly, lacking a clear physical interpretation. 

The key contribution of this work is two-fold. First, we provide a simple and intuitive geometrical interpretation for the apparent angle for non-vanishing wetting ridge, which we attribute to the rotation of the Neumann angle at the lubricant-droplet-gas contact line. Second, we derive a power series expansion of the apparent angle for small but non-vanishing wetting ridge. We validate all analytical expressions quantitatively against numerical results using Surface Evolver \cite{brakke1992surface}.

The paper is organised as follows. In section 2, we provide the mathematical derivation for the geometrical interpretation and an expansion of the apparent contact angle in terms of pressure differences between the lubricant, droplet and gas phases. In section 3, we list the essential physical assumptions and describe the computational method we employ to calculate droplet morphologies on liquid infused surfaces. We then compare the derived analytical expressions against the numerical results in section 4. Finally, we summarize our results in section 5.


\section{Theory}

We begin by discussing the limit of vanishing wetting ridge, where the lubricant pressure is much smaller than the pressure in both the droplet and gas phases. In this regime, the apparent angle $\theta_{\rm app}^{\rm S}$ (we use the superscript S to denote this regime) of a liquid droplet on LIS can be deduced from the force balance
\begin{align}
& \cos\theta_{\rm app}^{\rm S} = \frac{\gamma_{\rm gs}^{\rm eff} - \gamma_{\rm ws}^{\rm eff}}{\gamma_{\rm wg}}, \label{eq:Young}
\end{align}
where $\gamma_{\alpha\beta}$ is the surface tension between phases $\alpha$ and $\beta$ (droplet w, lubricant o and gas g), 
and the effective gas-solid and droplet-solid surface tensions capture the fact that the substrate is a composite of solid and lubricant. 
For simplicity, we will not resolve the details of the composite surface. Instead, we simply assume this gives rise to an
effective average surface tension $\gamma_{\alpha \rm s}^{\rm eff} = f_s \gamma_{\alpha \rm s} + (1-f_s) \gamma_{\alpha \rm o}$, with $f_s$ the fraction of the projected solid area exposed to the droplet or gas phase. We also note that when the lubricant is encapsulating the droplet, the suitable effective droplet-gas surface tension becomes the sum of droplet-lubricant and lubricant-gas surface tensions $\gamma_{\rm wg}^{\rm eff} = \gamma_{\rm wo} + \gamma_{\rm og}$ \cite{McHale_2019, Sadullah_2020_2}. Substituting the expressions for $\gamma_{\alpha \rm s}^{\rm eff}$ to Eq. \eqref{eq:Young} leads to
\begin{align}
\cos\theta_{\rm app}^{\rm S} = \frac{\gamma_{\rm og}}{\gamma_{\rm wg}}\cos\theta_{\rm og}^{\rm CB} - \frac{\gamma_{\rm ow}}{\gamma_{\rm wg}}\cos\theta_{\rm ow}^{\rm CB}, \label{eq:cosqappvm}
\end{align}
where the angle $\theta_{\rm o\alpha}^{\rm CB}$ models the wetting of the composite substrate in the spirit of the Cassie-Baxter model
\begin{align}
\cos\theta_{o\alpha}^{\rm CB} =f_s\cos\theta_{\rm o\alpha} + (1-f_s). \label{eq:cosqCB}
\end{align}
Here, $\theta_{\rm o\alpha}$ is defined as the Young's contact angle of the lubricant on the solid surface surrounded by the fluid $\alpha$ phase.

\begin{figure}[tb]
	\includegraphics[width=\linewidth,keepaspectratio]{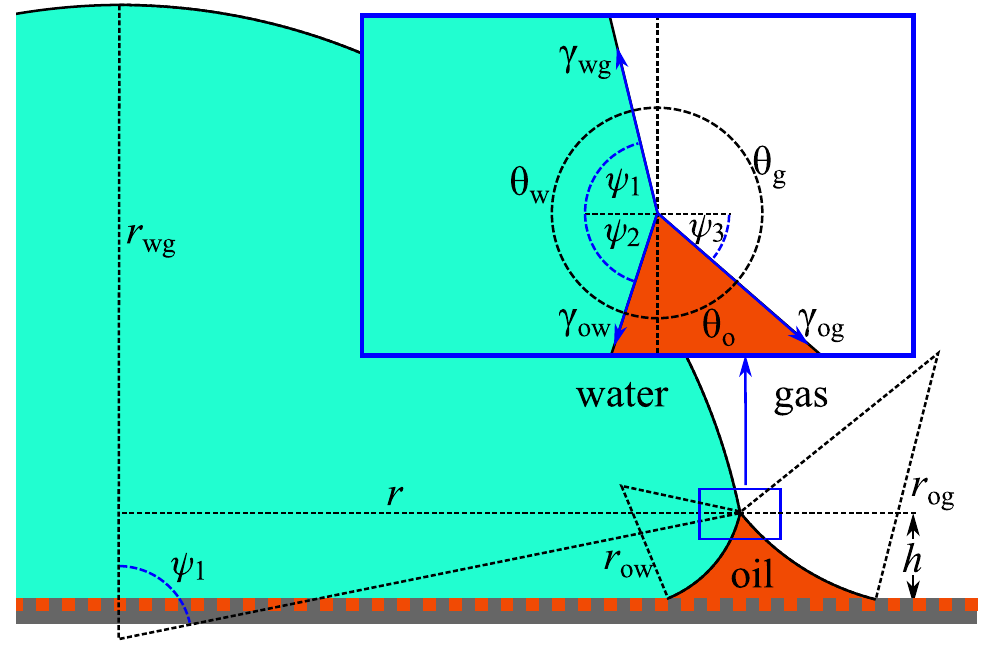}
	\caption{Sketch of the geometry of the oil ridge showing all curvature radii ($r_{\rm wg}$, $r_{\rm ow}$ and $r_{\rm og}$), the base radius of the lubricant-droplet-gas contact line ($r$), and the height of the wetting ridge ($h$). The inset illustrates the definition of the Neumann angles ($\theta_{\rm w}$, $\theta_{\rm g}$ and $\theta_{\rm o}$) and the geometrical angles $\psi_1$, $\psi_2$ and $\psi_3$ employed in Eqs. \ref{eq:psi1}, \ref{eq:psi2} and \ref{eq:psi3} respectively.
	}
	\label{fig1}
\end{figure}

With increasing lubricant pressure, more lubricant flows into the wetting ridge, and we have to account for the size and shape of the ridge. Following Semprebon et al. \cite{Semprebon2016b}, we can derive the following relation for the height of the wetting ridge
\begin{align}
h=& r_{\rm ow}\left[\cos\theta_{\rm ow}^{\rm CB}-\cos\left(\theta_{\rm w}-\theta_{\rm app}\right)\right] \nonumber \\
=&r_{\rm og}\left[\cos\theta_{\rm og}^{\rm CB}+\cos\left(\theta_{\rm app}+\theta_{\rm g}\right)\right]. \label{eq:fullsolutiongeom} 
\end{align}
In the above relation, both the lubricant-gas (og) and lubricant-droplet (ow) interfaces are assumed to be represented by circular arcs with radii $r_{\rm og}$ and $r_{\rm ow}$ respectively. This is valid when the in-plane curvature is much larger compared to the curvature in the azimuthal direction. Throughout this paper, this assumption is made in several places and we will systematically quantify its accuracy using comparisons against our numerical results. In Eq. \eqref{eq:fullsolutiongeom}, we have also dropped the superscript for the apparent angle to denote the wetting ridge size is no longer negligible and used the Neumann angles, $\theta_{\rm w}$, $\theta_{\rm g}$, $\theta_{\rm o}$ ($\theta_{\rm w}+\theta_{\rm g}+\theta_{\rm o}=2\pi$), as illustrated in Fig. \ref{fig1}. The Neumann angles are related to the surface tensions via the relation 
\begin{align}
\frac{\sin\theta_{\rm g}}{\gamma_{\rm ow}}=\frac{\sin\theta_{\rm o}}{\gamma_{\rm wg}}=\frac{\sin\theta_{\rm w}}{\gamma_{\rm og}}. \label{eq:neumann}
\end{align}

Next, we consider the cyclic relation for the droplet, lubricant and gas pressures given by
\begin{align}
\Delta P_{\rm wg}=\Delta P_{\rm wo}+\Delta P_{\rm og}. 
\label{eq:pressurepermutation}
\end{align}
Let us discuss the 2D case first, for which the azimuthal curvature is absent and all the fluid-fluid interfaces are exactly represented by circular arcs. In this case we set $\Delta P_{\rm wg}=\gamma_{\rm wg}/r_{\rm wg}$, $\Delta P_{\rm og}=-\gamma_{\rm og}/r_{\rm og}$ and $\Delta P_{\rm wo}=\gamma_{\rm wo}/r_{\rm wo}$ to write
\begin{align}
\frac{\gamma_{\rm wg}}{r_{\rm wg}}=\frac{\gamma_{\rm ow}}{r_{\rm ow}}-\frac{\gamma_{\rm og}}{r_{\rm og}}, \label{eq:pressurepermutation2D}
\end{align}
which, with a straightforward manipulation, leads to
\begin{align}
\frac{r_{\rm og}}{r_{\rm ow}}=\frac{\gamma_{\rm og}}{\gamma_{\rm ow}}\left(1-\frac{1}{\alpha}\frac{\Delta P_{\rm wg}}{\Delta P_{\rm og}}\right). \label{eq:rratio}
\end{align}
Note that we have introduced the parameter $\alpha$ that takes a value of $\alpha=1$ in 2D.

For the 3D case, following the geometry in Fig. \ref{fig1}, the Laplace pressures at the lubricant-droplet-gas contact line are
\begin{align}
\Delta P_{\rm wg}&=\frac{2\gamma_{\rm wg}}{r_{\rm wg}}=\gamma_{\rm wg}\left(\frac{1}{r_{\rm wg}}+\frac{\sin\psi_1}{r}  \right), \label{eq:psi1}\\
\Delta P_{\rm wo}&\simeq\gamma_{\rm wo}\left(\frac{1}{r_{\rm ow}}+\frac{\sin\psi_2}{r}  \right), \label{eq:psi2}\\
\Delta P_{\rm og}&\simeq\gamma_{\rm wo}\left(-\frac{1}{r_{\rm og}}+\frac{\sin\psi_3}{r}  \right), \label{eq:psi3}
\end{align}
where $\Psi_1=\theta_{\rm app}$ and we have introduced the radius of the lubricant-droplet-gas contact line $r=r_{\rm wg}\sin\psi_1$. Eq. \eqref{eq:psi1} is exact as the droplet-gas interface retains the shape of a spherical cap, while Eqs. \eqref{eq:psi2} and \eqref{eq:psi3} are approximated where we assume the in-plane shape of the interfaces can be described as circular arcs. In general, the lubricant-droplet and lubricant-gas interfaces will assume more complex Delaunay surfaces, with non constant curvature along the radial cross section. Summing the azimuthal contributions (second terms on the right hand side) in all three equations, we obtain
\begin{align}
\frac{1}{r}\left(\gamma_{\rm wg}\sin\psi_1 + \gamma_{\rm ow}\sin\psi_2 + \gamma_{\rm og}\sin\psi_3  \right)=0, \label{eq:sumazimutal}
\end{align}
where the term in bracket vanishes as it represents the force balance in the vertical direction at the triple point. Substituting this cancellation to the pressure cyclic relation, we find the same relation as in Eqs. \eqref{eq:pressurepermutation2D} and \eqref{eq:rratio}, except that now $\alpha = 2$ for the 3D case. As such, both the 2D and 3D cases can be treated in the same way from this point on. Introducing Eq. \eqref{eq:rratio} into Eq. \eqref{eq:fullsolutiongeom} we obtain 
\begin{align}
\frac{\sin\theta_{\rm g}\left[\cos\theta_{\rm ow}^{\rm CB}-\cos\left(\theta_{\rm w}-\theta_{\rm app}\right)\right]}{\sin\theta_{\rm w}\left[\cos\theta_{\rm og}^{\rm CB}+\cos\left(\theta_{\rm app}+\theta_{\rm g}\right)\right]}=\left(1-\frac{1}{\alpha}\frac{\Delta P_{\rm wg}}{\Delta P_{\rm og}}\right). \label{eq:fullsolution}
\end{align}


\subsection{Geometric interpretation}
In this subsection, we can recast Eq. \eqref{eq:fullsolution} to provide a simple geometric interpretation for the apparent contact angle. Expanding Eq. \eqref{eq:fullsolution}, it can be shown that it leads to 
\begin{align}
-[\sin \theta_{\rm g} \cos \theta_{\rm w} + \sin \theta_{\rm w} \cos \theta_{\rm g}] \cos \theta_{\rm app} = \nonumber \\
\sin \theta_{\rm w} \cos \theta_{\rm og}^{\rm CB} - \sin \theta_{\rm g} \cos \theta_{\rm ow}^{\rm CB} \\
- \sin \theta_{\rm w} [\cos \theta_{\rm og}^{\rm CB} + \cos (\theta_{\rm app} + \theta_{\rm g})] \frac{1}{\alpha}\frac{\Delta P_{\rm wg}}{\Delta P_{\rm og}}. \nonumber
\end{align}

Furthermore, recognising that $\sin \theta_{\rm g} \cos \theta_{\rm w} + \sin \theta_{\rm w} \cos \theta_{\rm g} = \sin (\theta_{\rm g} + \theta_{\rm w}) = -\sin \theta_{\rm o}$, $\sin \theta_{\rm w}/\sin \theta_{\rm 0} = \gamma_{\rm og} / \gamma_{\rm wg}$, $\sin \theta_{\rm g}/\sin \theta_{\rm 0} = \gamma_{\rm ow}/\gamma_{\rm wg}$, and $\sin \theta_{\rm w}/\sin \theta_{\rm 0} \times  \Delta P_{\rm wg}/\alpha \Delta P_{\rm og} = - r_{\rm og}/r_{\rm wg}$, the above equation can be simplified to
\begin{eqnarray}
\cos\theta_{\rm app} &=& \frac{\gamma_{\rm og}}{\gamma_{\rm wg}}\cos\theta_{\rm og}^{\rm CB} - \frac{\gamma_{\rm ow}}{\gamma_{\rm wg}}\cos\theta_{\rm ow}^{\rm CB} \\
&& + \frac{r_{\rm og}}{r_{\rm wg}} [\cos \theta_{\rm og}^{\rm CB} + \cos (\theta_{\rm app} + \theta_{\rm g})]. \nonumber
\end{eqnarray}

The first two terms on the right hand side are in fact our definition for $\cos\theta_{\rm app}^{\rm S}$ in Eq.~\eqref{eq:cosqappvm}. The numerator of the final term is nothing else than Eq.~\eqref{eq:fullsolutiongeom} and the final term can be simplified to $h/r_{\rm wg}$. Therefore
\begin{align}
\cos\theta_{\rm app}=\cos\theta_{\rm app}^{\rm S}+\frac{h}{r_{\rm wg}}, \label{eq:geometric}
\end{align}
and
\begin{align}
 \Delta \theta_{\rm app} = \theta_{\rm app} - \theta_{\rm app}^{\rm S}\simeq\frac{h}{r_{\rm wg} \sin \theta_{\rm app}} = \frac{h}{r}. \label{eq:geometrictheta}
\end{align}
 
\begin{figure}[tb]
	\includegraphics[width=0.8\linewidth,keepaspectratio]{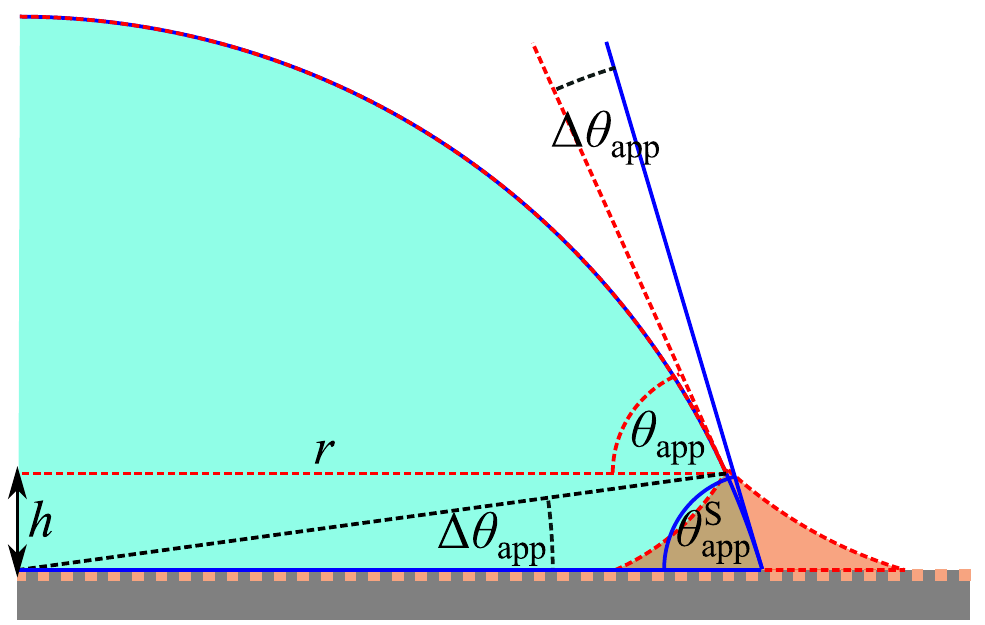}	
	\caption{Geometric illustration of the rotation of the apparent angle on LIS by $\Delta \theta_{\rm app}$. The apparent angles $\theta_{\rm app}^{\rm S}$ and $\theta_{\rm app}$ represent the vanishing and finite wetting ridge cases.
	}
	\label{fig2}
\end{figure}

This states that, with increasing wetting ridge height, the Neumann triangle and hence the apparent angle rotates from the value at the vanishing ridge case by the ratio between the ridge height and the radius of the lubricant-droplet-gas contact line, $h/r$. Thus, increasing wetting ridge height reduces the apparent contact angle. This rotation is illustrated geometrically in Fig. \ref{fig2}.

\subsection{Power Series Expansion}
While the geometric interpretation in Eqs. \eqref{eq:geometric} and \eqref{eq:geometrictheta} are simple, intuitive and self-consistent, they are not a priori predictive because we know neither the apparent angle nor the wetting ridge height. However, Eq. \eqref{eq:fullsolution} can also be recast in a quadratic equation for $\cos\theta_{\rm app}$ given by
\begin{align}
\cos\theta_{\rm app} =\frac{AC+B\sqrt{A^2+B^2-C^2}}{A^2+B^2}, \label{eq:cosqappFS}
\end{align}
where
\begin{align}
A&=\sin\theta_{\rm g}\cos\theta_{\rm w}+\left(1-\frac{1}{\alpha}\frac{\Delta P_{\rm wg}}{\Delta P_{\rm og}}\right)\sin\theta_{\rm w}\cos\theta_{\rm g}, \label{eq:A}\\
B&=\sin\theta_{\rm g}\sin\theta_{\rm w}\frac{1}{\alpha}\frac{\Delta P_{\rm wg}}{\Delta P_{\rm og}}, \label{eq:B}\\
C&=\sin\theta_{\rm g}\cos\theta_{\rm ow}^{\rm CB}-\left(1-\frac{1}{\alpha}\frac{\Delta P_{\rm wg}}{\Delta P_{\rm og}}\right)\sin\theta_{\rm w}\cos\theta_{\rm og}^{\rm CB}. \label{eq:C}
\end{align}
In this form, once the pressure ratio $\Delta P_{\rm wg}/\Delta P_{\rm og}$ is known, the apparent contact angle can be computed since all the other quantities correspond to material parameters. Eq. \eqref{eq:cosqappFS} is exact in 2D, but it is only an approximation in 3D due to the circular arc approximations for the in-plane curvatures of the lubricant-droplet and lubricant-gas interfaces.

To provide additional insights, we can further expand Eq. \eqref{eq:cosqappFS} using the following small parameter 
\begin{align}
\xi=-\frac{\gamma_{\rm og}}{\gamma_{\rm wg}}\frac{1}{\alpha}\frac{\Delta P_{\rm wg}}{\Delta P_{\rm og}}= \frac{r_{\rm og}}{r_{\rm wg}}. \label{eq:xi}
\end{align}
Performing the expansion, 
\begin{align}
& A = -\sin\theta_{\rm o} [1-\xi\cos\theta_{\rm g}],\label{eq:Anew}\\
& B = -\xi \sin\theta_{\rm o} \sin\theta_{\rm g},\label{eq:Bnew} \\
& C = - \sin\theta_{\rm o} [\cos\theta_{\rm app}^{\rm S}+\xi\cos\theta_{\rm og}^{\rm CB}],\label{eq:Cnew}
\end{align}
and we find
\begin{align}
\cos\theta_{\rm app} \simeq & \cos\theta_{\rm app}^{\rm S}+\left[\cos(\theta_{\rm app}^{\rm S}+\theta_{\rm g})+\cos\theta_{\rm og}^{\rm CB}\right]\xi \nonumber \\
& +\mathcal{O}(\xi^2). \label{eq:linear}
\end{align} 
The linear order term of this expansion is more compact compared to that presented by Semprebon et al. \cite{Semprebon2016b}. It is also pleasing to see that the geometric interpretation in Eq. \eqref{eq:geometric} reduces to  Eq. \eqref{eq:linear} in the limit of small ridges, since $\theta_{\rm app} \rightarrow \theta_{\rm app}^{\mathrm S}$. Furthermore, for completeness, we can also carry out the expansion to quadratic order and the result reads
\begin{align}
\cos\theta_{\rm app} \simeq & \cos\theta_{\rm app}^{\rm S}+\Lambda_1\xi+\Lambda_2\xi^2 +\mathcal{O}(\xi^3) \label{eq:secondorder}
\end{align} 
where
\begin{align}
\Lambda_1=\left[\cos(\theta_{\rm app}^{\rm S}+\theta_{\rm g})+\cos\theta_{\rm og}^{\rm CB}\right], \label{eq:secondorder1} \\ 
\Lambda_2=\left[\cos(\theta_{\rm app}^{\rm S}+\theta_{\rm g})+\cos\theta_{\rm og}^{\rm CB}\right]\frac{\sin(\theta_{\rm app}^{\rm S}+\theta_{\rm g})}{ \sin\theta_{\rm app}^{\rm S}}. \label{eq:secondorder2}
\end{align}


\section{Numerical implementation}

To numerically compute the equilibrium shapes of the droplet and lubricant interfaces, we employ a finite element approach based on the free software Surface Evolver \cite{brakke1992surface} to minimise the total energy of the system, which is given by
\begin{align}
E=\sum_{\alpha\neq\beta}\gamma_{\alpha\beta}A_{\alpha\beta}+\sum_\alpha\gamma_{\alpha {\rm s}} A_{\alpha {\rm s}} + \Delta P_{\rm wg} V_{\rm w} + \Delta P_{\rm og} V_{\rm o}.
\label{eq:energy}
\end{align}
Here $\gamma_{\alpha\beta}$ and $A_{\alpha\beta}$ denote the surface tension and interfacial area between any two fluid phases (water $\rm w$, lubricant $\rm o$ and gas $\rm g$), while $\gamma_{\alpha {\rm s}}$  and $A_{\alpha s}$ correspond to those between any fluid phase $\alpha$ and the solid surface. In this work, we use volume ensembles for the droplet and lubricant phases such that the pressure differences, $\Delta P_{\rm wg}$ and $\Delta P_{\rm og}$, act as Lagrange multipliers to the volume of the liquid droplet $V_{\rm w}$ and the lubricant wetting ridge $V_{\rm o}$. 

Both 2D and rotationally symmetric 3D systems are modelled by representing the interfaces through geometric elements such as nodes and edges. To optimise the convergence the algorithm, at the beginning of the calculation and whenever one control parameter is modified, we typically coarsen the mesh to allow efficient displacement of the geometric elements. This is then followed by a progressive mesh refinement to obtain more accurate shapes. For a minimised configuration, we then extract all relevant geometric parameters, including the apparent angle at the lubricant-droplet-gas contact, the curvature radii of various fluid-fluid interfaces, the ridge height, and the Laplace pressures. Different values of pressure ratios are obtained by varying the target value for the constraint on the lubricant volume. 

\section{Validation}.

We now provide a validation of the model for the 2D case and a benchmark for the accuracy of the approximations in the 3D case for a wide range of surface tension parameters.

\begin{figure}[tb]
	\includegraphics[width=0.8\linewidth,keepaspectratio]{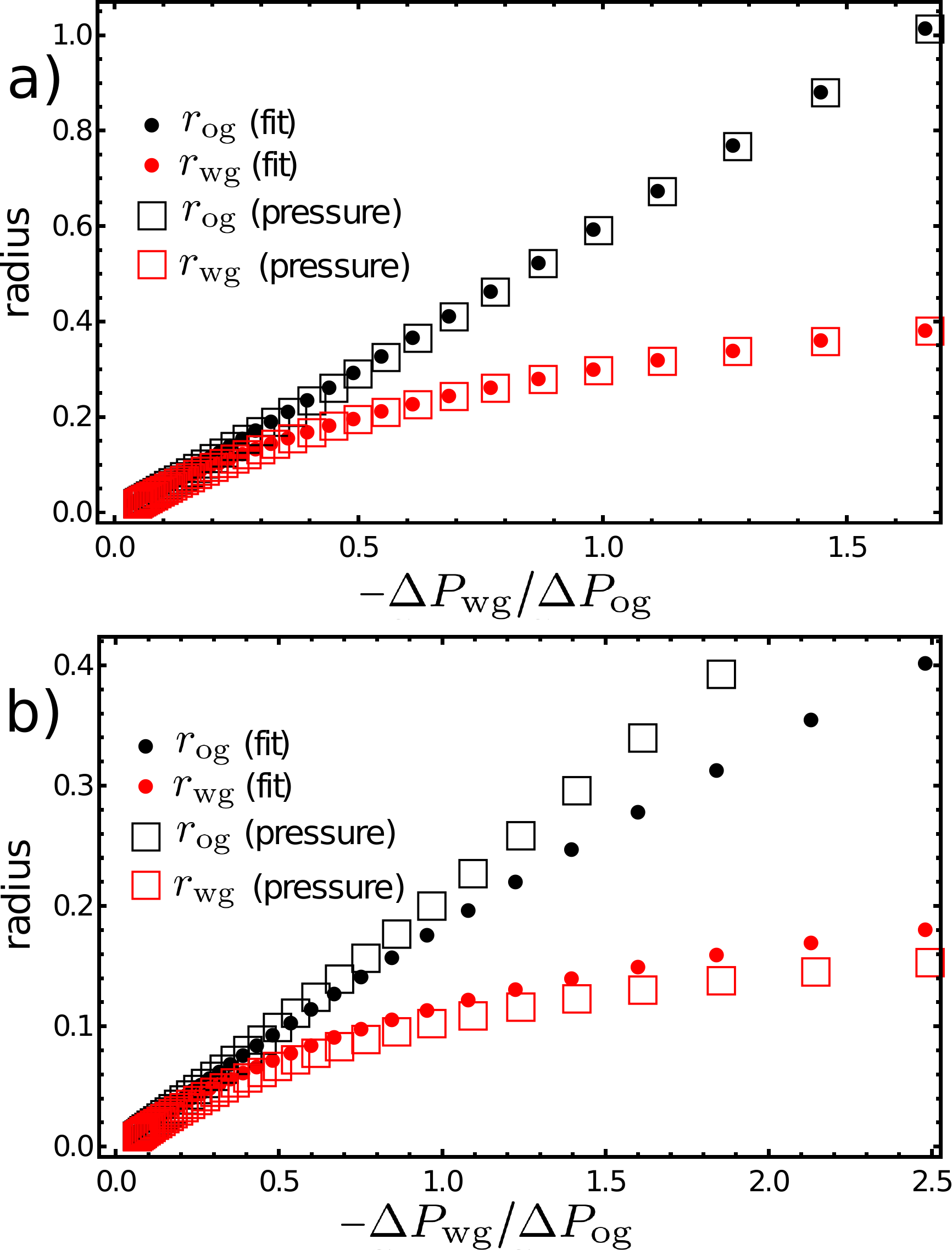}
	\caption{The lubricant-droplet and lubricant-gas curvature radii as function of the pressure ratio $-\Delta P_{\rm wg}/\Delta P_{\rm og}$ for the 2D (panel a) and 3D (panel b) cases. Two measures for the curvature radii are presented, by inverting the Laplace pressures measured and through fitting the interface shapes with circular arcs. The parameters used here are $\theta_{\rm o}=40^\circ$, $\theta_{\rm w}=160^\circ$, $\theta_{\rm g}=160^\circ$, $\theta_{\rm ow}^{\rm CB}=15^\circ$ and $\theta_{\rm og}^{\rm CB}=15^\circ$.
	}
	\label{fig3}
\end{figure}

We begin by assessing the accuracy of the circular arc approximation for the in-plane lubricant-droplet and lubricant-gas curvatures as we vary the wetting ridge height. The ridge height depends on the pressure ratio $-\Delta P_{\rm wg}/\Delta P_{\rm og}$. In Fig. \ref{fig3} we report a comparison between two different estimates for the curvature radii when the following parameters are used: $\theta_{\rm o}=40^\circ$, $\theta_{\rm w}=\theta_{\rm g}=160^\circ$, and $\theta_{\rm ow}^{\rm CB}=\theta_{\rm og}^{\rm CB}=15^\circ$. The first estimate fits the numerically computed shapes of the interfaces with circular arcs using three points (the two end points and the middle point of the interfaces). The second estimate is obtained by inverting the Laplace pressures, which can be extracted from Surface Evolver as the Lagrange multipliers of the volume constraints enforced. Since we have constrained the droplet and lubricant volumes,  the corresponding Lagrange multipliers describe the Laplace pressures of the droplet-gas and lubricant-gas interfaces. Hence, $r_{\rm og}=-\gamma_{\rm og}/\Delta P_{\rm og}$ and $r_{\rm ow}=\gamma_{\rm ow}/(\Delta P_{\rm wg}-\Delta P_{\rm og})$. As can be observed in Fig. \ref{fig3}, for the 2D case the two estimates coincide with high accuracy. For the 3D case, assuming the fit (first method) estimate represents a more accurate representation of the in-plane curvature, we can take its discrepancy with the pressure (second method) estimate to evaluate the effects of neglecting the azimuthal curvatures. The pressure approach overestimates the in-plane radius of curvature for the lubricant-gas interface, and underestimates the in-plane curvature for the lubricant-water interface. Overall, the circular arc approximation is a better assumption for the lubricant-droplet than for the lubricant-gas interface. As expected, the different estimates converge as $-\Delta P_{\rm wg}/\Delta P_{\rm og}$ goes to zero, when we approach the vanishing wetting ridge limit.

\begin{figure}[tb]
	\includegraphics[width=0.8\linewidth,keepaspectratio]{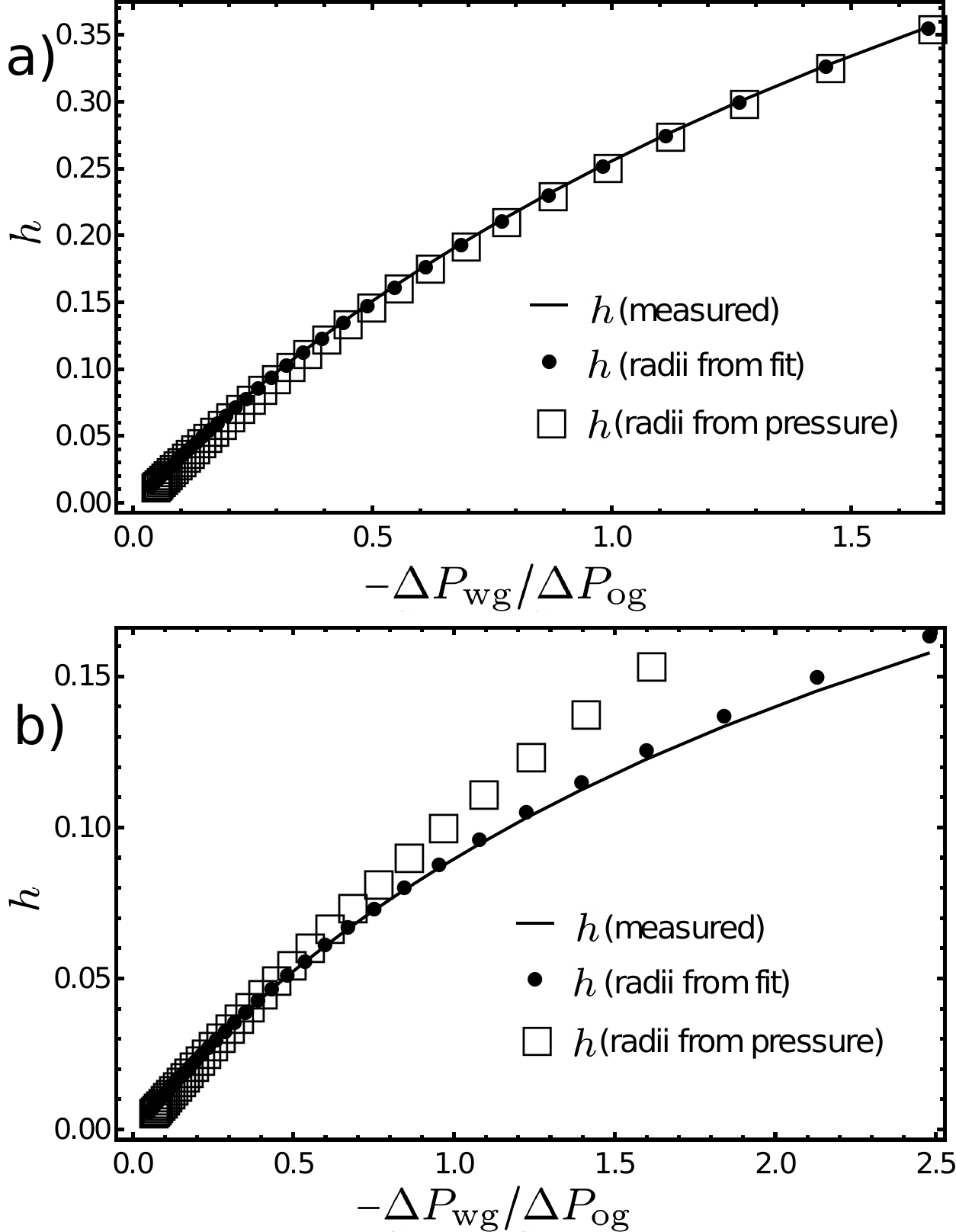}
	\caption{Comparison of the measured lubricant ridge $h$ with those computed according to Eq. \eqref{eq:fullsolutiongeom}. The radii $r_{\rm ow}$ and $r_{\rm og}$ can obtained by fitting the interfaces with circular arcs or by inverting the Laplace pressures, as shown in Fig. \ref{fig3}. The parameters used here are $\theta_{\rm o}=40^\circ$, $\theta_{\rm w}=160^\circ$, $\theta_{\rm g}=160^\circ$, $\theta_{\rm ow}^{\rm CB}=15^\circ$ and $\theta_{\rm og}^{\rm CB}=15^\circ$. Panel a) 2D; panel b) 3D. 
	}
	\label{fig4}
\end{figure}

To further assess the impact of the radius of curvature estimates for the prediction of the apparent contact angles, we report in Fig. \ref{fig4} the comparison between the measured lubricant ridge height $h$, and those calculated based on Eq. \eqref{eq:fullsolutiongeom}, where $r_{\rm ow}$ and $r_{\rm og}$  are either fitted from the interface profiles or estimated via the Laplace pressures. For the 2D case, the correspondence is accurate in both cases, while for the 3D case, the values calculated using fitted $r_{\rm ow}$ and $r_{\rm og}$ agree better than those from Laplace pressures. This is a consequence of neglecting the azimuthal curvature in 3D that becomes more important for larger $-\Delta P_{\rm wg}/\Delta P_{\rm og}$.

\begin{figure}[tb]
	\includegraphics[width=0.8\linewidth,keepaspectratio]{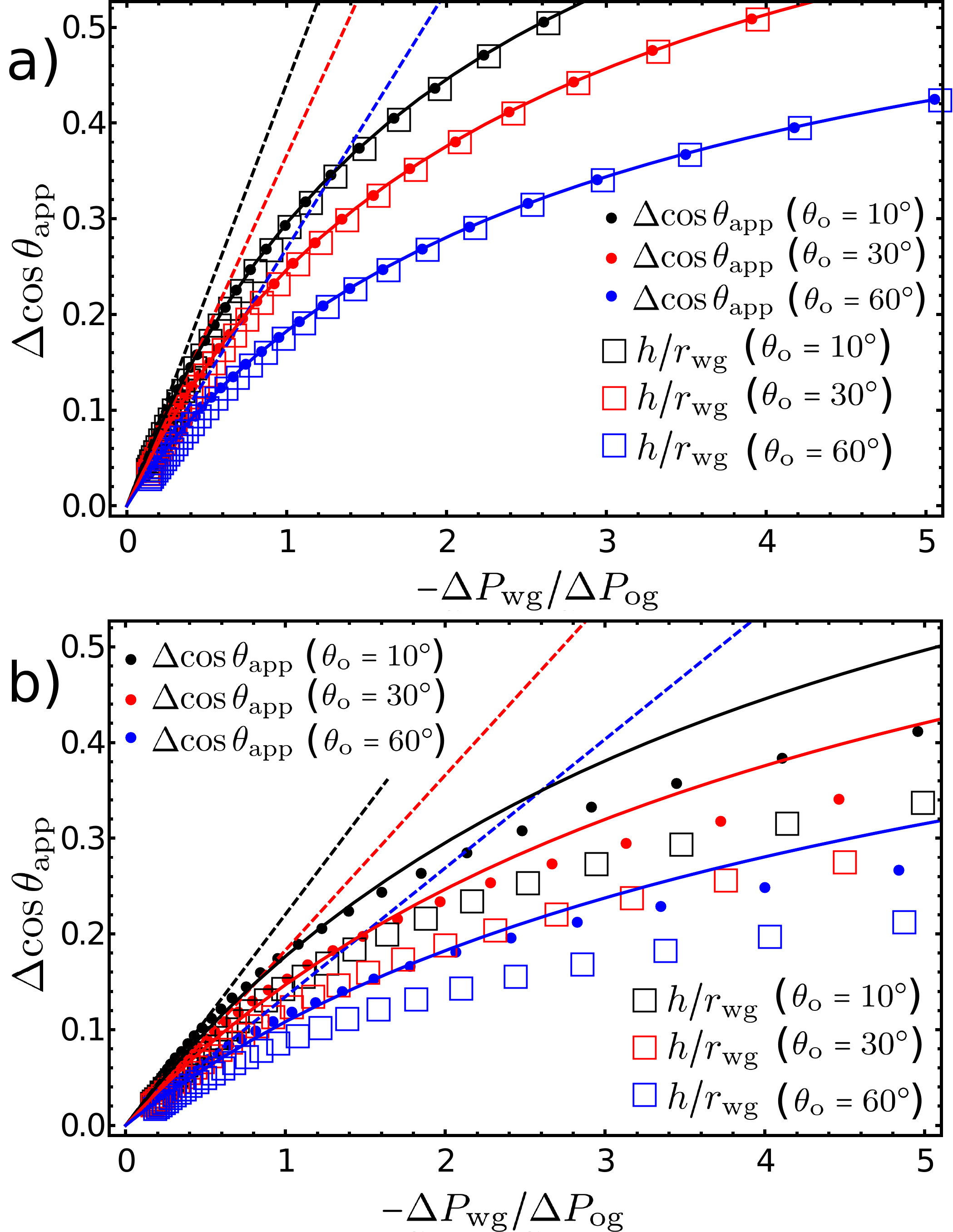}
	\caption{Deviation of the apparent contact angle from the vanishing meniscus ridge value as a function of the pressure ratio $-\Delta P_{\rm wg}/\Delta P_{\rm og}$ for the symmetric scenario where $\theta_{\rm w}=\theta_{\rm g}$, and $\theta_{\rm ow}^{\rm CB}=\theta_{\rm og}^{\rm CB}=15^\circ$. Here, three values of lubricant Neumann angle are presented, corresponding to $\theta_{\rm o}=10^\circ$, $\theta_{\rm o}=30^\circ$ and $\theta_{\rm o}=60^\circ$. The filled points are the measured $\Delta \cos\theta_{\rm app}$; the void squares are the measured $h/r_{\rm wg}$; the dashed lines represent the linear expansion in Eq. \eqref{eq:linear}; and the continuous lines correspond to Eq. \eqref{eq:cosqappFS}. Panel a) 2D; panel b) 3D.
	}
	\label{fig5}
\end{figure}

We now focus our attention on the accuracy of the various apparent contact angle descriptions as a function of the pressure ratio $-\Delta P_{\rm wg}/\Delta P_{\rm og}$. We first consider a symmetric scenario in Fig. \ref{fig5}, setting $\theta_{\rm w}=\theta_{\rm g}$ and varying $\theta_{\rm o}=10^\circ$, $\theta_{\rm o}=30^\circ$ and $\theta_{\rm o}=60^\circ$. The two wetting angles are $\theta_{\rm ow}^{\rm CB}=\theta_{\rm og}^{\rm CB}=15^\circ$. For these parameters the apparent contact angle in the limit of the vanishing ridge is the same, $\theta_{\rm app}^{\rm S}=90^\circ$, but its variation with $-\Delta P_{\rm wg}/\Delta P_{\rm og}$ depends on the value of $\theta_{\rm o}$. We can observe that, in 2D, the values of $\Delta \cos\theta_{\rm app}$ coincide exactly with the measure of $h/r_{\rm wg}$ for the whole range of pressures. Furthermore, these data points are also accurately captured by the continuous lines representing Eq. \eqref{eq:cosqappFS}. In 3D, we observe that $h/r_{\rm wg}$ systematically underestimates the measured values of $\Delta \cos\theta_{\rm app}$, while Eq. \eqref{eq:cosqappFS} overestimates it. 
We can attribute the discrepancy between Eq. \eqref{eq:geometric} (the geometric interpretation) and Eq. \eqref{eq:cosqappFS} to the different ways in which the azimuthal curvature affects the model assumptions. In both models, neglecting azimuthal curvature influences the accuracy of the pressure ratio as the control parameter in Eq. \eqref{eq:fullsolution}. In addition, for the geometric interpretation, the  azimuthal curvature assumption also enters the determination of the wetting ridge height $h$, as analysed in Fig. \ref{fig4}.
In the limit of vanishing pressure ratios, all curves converge. Any small deviation observed as $-\Delta P_{\rm wg}/\Delta P_{\rm og} \rightarrow 0$ between the measured $\cos\theta_{\rm app}$ and the analytical models is mainly due to numerical limitations and inaccuracies. As the wetting ridge becomes smaller, we require increasingly finer refinements in Surface Evolver. It is worth noting that the measured and predicted contact angles are in agreement to within a degree in this limit.

\begin{figure}[tb]
	\includegraphics[width=0.8\linewidth,keepaspectratio]{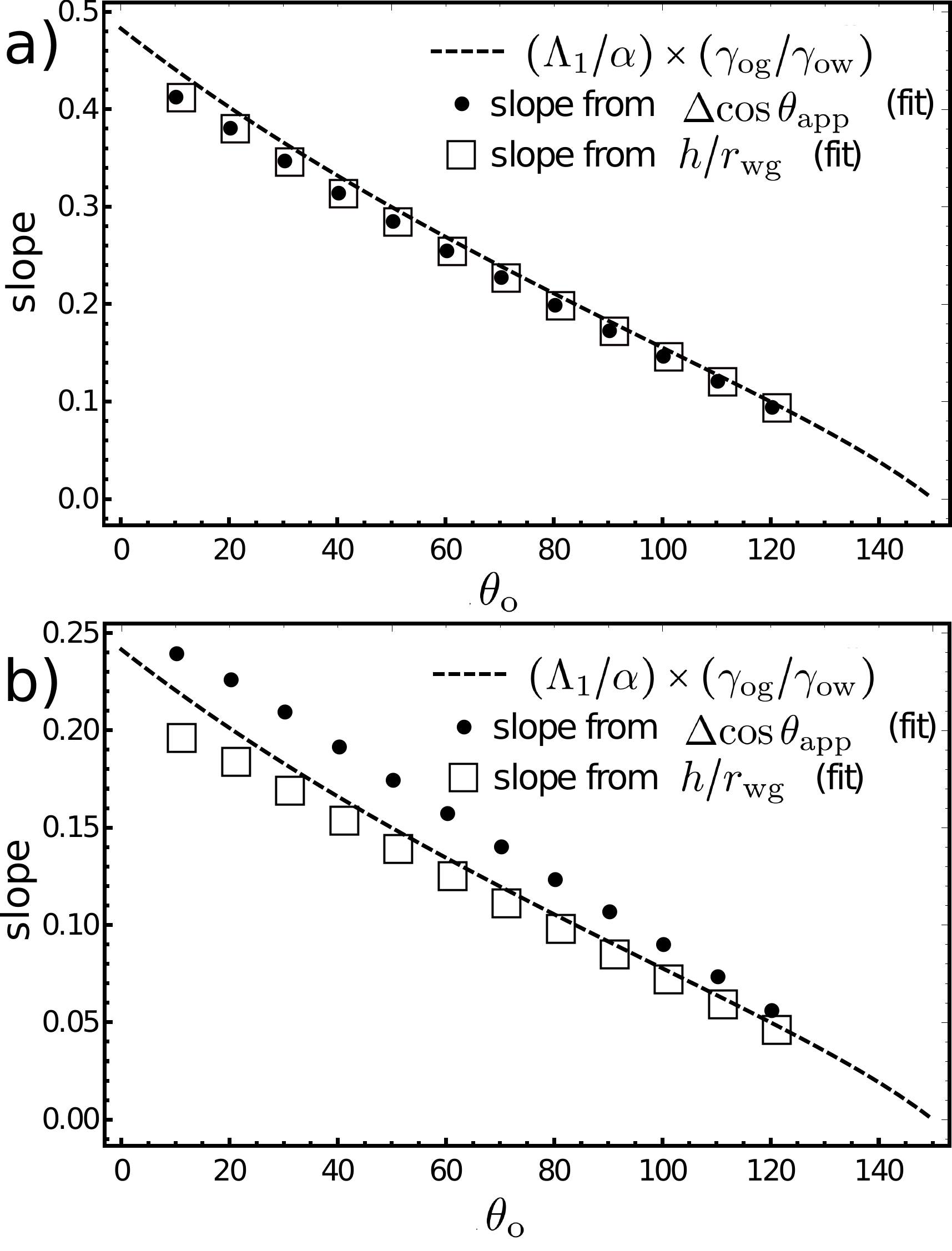}
	\caption{The linear slopes of $\cos\theta_{\rm app}$ (filled circles) and $h/r_{\rm wg}$ (empty squares) vs $-\Delta P_{\rm wg}/\Delta P_{\rm og}$ as we vary $\theta_{\rm o}$. The rest of the simulation parameters are the same as in Fig. \ref{fig5}. The dashed line represents the linear equation derived in Eq. \eqref{eq:linear}. Panel a) 2D; panel b) 3D.
	}
	\label{fig6}
\end{figure}

Focussing on the small but non-zero $-\Delta P_{\rm wg}/\Delta P_{\rm og}$ limit, we can measure the slopes of $\cos\theta_{\rm app}$ and $h/r_{\rm wg}$ vs $-\Delta P_{\rm wg}/\Delta P_{\rm og}$, and compare these against the prediction from the linear expansion in Eq. \eqref{eq:linear}. Fig. \ref{fig6} shows the results for a wide range of $\theta_{\rm o}$.  All the other parameters are the same as in Fig. \ref{fig5}. Overall, the agreement is excellent both in 2D and 3D with the trends as a function of $\theta_{\rm o}$ clearly reproduced. We do find the discrepancy is larger in 3D because the 3D model is more complex and hence the energy minimization is more difficult to converge in Surface Evolver. 

\begin{figure}[tb]
	\includegraphics[width=0.8\linewidth,keepaspectratio]{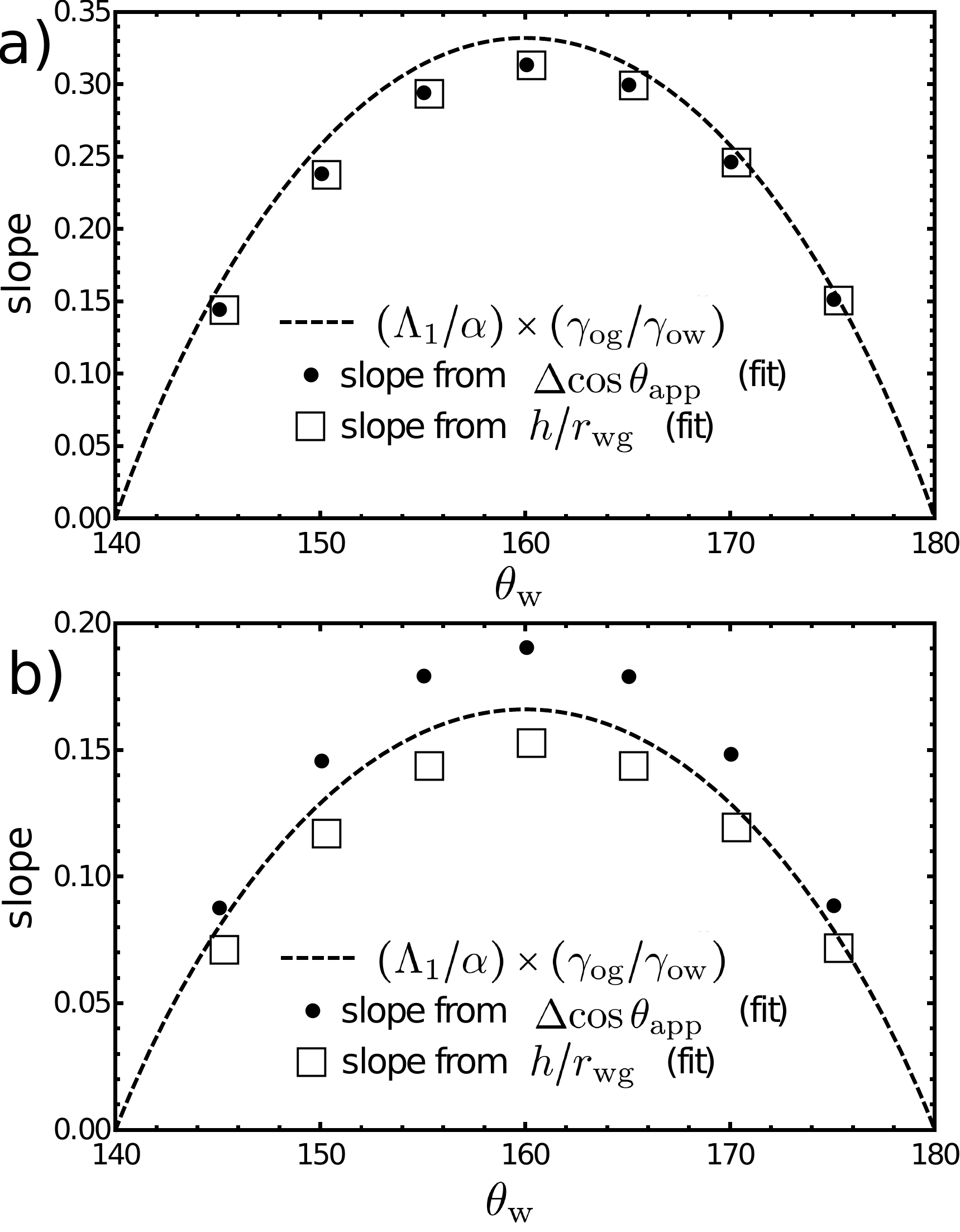}
	\caption{
	The linear slopes of $\cos\theta_{\rm app}$ (filled circles) and $h/r_{\rm wg}$ (empty squares) vs $-\Delta P_{\rm wg}/\Delta P_{\rm og}$ as we vary $\theta_{\rm w}$. The other parameters used here are $\theta_{\rm o}=40^\circ$, and $\theta_{\rm ow}^{\rm CB}=\theta_{\rm og}^{\rm CB}=15^\circ$. The dashed line represents the linear equation derived in Eq. \eqref{eq:linear}. Panel a) 2D; panel b) 3D.
		}
	\label{fig7}
\end{figure}

To complete our validation we also consider asymmetric cases, where we fix the lubricant Neumann angle at  $\theta_{\rm o}=40^\circ$ and vary the droplet ($\theta_{\rm w}$) and gas ($\theta_{\rm g}$) Neumann angles. The wetting angles are again set constant at $\theta_{\rm ow}^{\rm CB}=\theta_{\rm og}^{\rm CB}=15^\circ$.
In this case $\cos\theta_{\rm app}^{\rm S}$ varies.
As before, we compute the linear slopes of $\cos\theta_{\rm app}$ and $h/r_{\rm wg}$ vs $-\Delta P_{\rm wg}/\Delta P_{\rm og}$ for varying $\theta_{\rm w}$. Here, the physically meaningful values of $\theta_{\rm w}$ are restricted to the interval $[180^\circ-\theta_{\rm o},180^\circ]$. As shown in Fig. \ref{fig7}, the maximum in the slopes occur when $\theta_{\rm w} = \theta_{\rm g}$, and they symmetrically decrease and go to zero as we approach the limits of the physical interval for $\theta_{\rm w}$. As in the symmetric scenario in Fig. \ref{fig6}, the agreement between the measured and predicted slopes are excellent. Minor deviations can be seen for the 2D results, which we can attribute to the inaccuracies in our fitting procedures. The larger deviations in 3D are again due to the more difficult convergence of the model in Surface Evolver. It is worth emphasising that, when comparing the values of the contact angles, the predicted and measured values all agree within a degree.

\section{Conclusions}
In this work we have provided several new insights on the effect of finite wetting ridge size on the apparent contact angle of droplets on LIS. Importantly, we have deduced a simple and intuitive geometrical interpretation that the reduction of the apparent angle with increasing ridge size is due to the rotation of the Neumann triangle at the lubricant-droplet-gas contact line. This rotation is given by the ratio of the ridge height and the lubricant-droplet-gas contact line radius. Comparing the analytical predictions against numerical data from Surface Evolver, we find this interpretation is highly accurate in 2D across the whole range of ridge height, while in 3D this approximation becomes poorer with increasing ridge height. The latter is due to our assumption to ignore the azimuthal curvature for the lubricant-droplet and lubricant-gas interfaces. In addition, we have also performed power series expansion of the apparent angle in terms of a small parameter $\xi$ that is related to the pressure differences between the lubricant, droplet and gas phases. The expressions derived here are much simpler, and yet equally accurate, compared to those previously described in the literature.

\begin{acknowledgments}
CS acknowledges support from Northumbria University through the
Vice-Chancellor’s Fellowship Programme and EPSRC (EP/S036857/1) for funding. MSS is supported by funding from King Abdullah University of Science and Technology (KAUST) under the award number BAS/1/1070-01-01. HK acknowledges funding from EPSRC (EP/V034154/1).
\end{acknowledgments}


\bibliographystyle{apsrev4-1}
\bibliography{biblio}

\end{document}